\documentclass[aps,preprint,preprintnumbers,nofootinbib,showpacs]{revtex4}
\usepackage{graphicx}
\begin{document}
\title{$p$-brane production in Fat brane or Universal extra dimension scenario}
\author{Kingman Cheung}
\email[Email:]{cheung@phys.cts.nthu.edu.tw}
\affiliation{National Center for Theoretical Sciences, National Tsing Hua
University, Hsinchu, Taiwan, R.O.C.}
\author{Chung-Hsien Chou}
\email[Email:]{chouch@phys.sinica.edu.tw}
\affiliation{
Institute of Physics, Academia Sinica, Taipei, Taiwan 115, R.O.C.}
\date{\today}

\begin{abstract}
In models of large extra dimensions, the fundamental Planck scale can be
as low as TeV.  Thus, in hadronic collisions interesting objects like
black holes, string balls, or $p$-branes can be produced.
In scenarios of fat brane or universal extra dimensions,
the SM particles are allowed to propagate in the extra spatial dimensions,
which leads to the enhancement of production cross sections of
black holes and $p$-branes.
Especially, the ratio of $p$-brane cross section to the black hole cross
section increases substantially, in comparison with the original confined
scenario.  The ratio can be as large as $105$ (for
the case $n=7,m=5=p=r=k$).
\end{abstract}
\pacs{}
\preprint{NSC-NCTS-020524}
\maketitle

\section{Introduction}

Trans-Planckian objects including black holes (BH) recently receive a lot of
attentions in models of large extra dimensions.  In an attractive
model of large extra dimensions or TeV quantum gravity (ADD model)
\cite{arkani}, the fundamental Planck scale can be as low as a few TeV.
This is made possible by confining the SM particles on a
3-brane (using the idea of D-branes in Type I or II string theory),
while gravity is free to propagate in all dimensions.
The observed Planck scale ($\sim 10^{19}$ GeV) is then a derived quantity.
Extensive phenomenological studies have been carried out in past few years.
Signatures for the ADD model can be divided into two categories:
sub-Planckian and trans-Planckian.
The former is the one that was studied extensively, while
the latter just recently receives more attentions, especially, black hole
production in hadronic collisions.

In models of large extra dimensions, the properties of black holes
are modified and interesting signatures emerge
\cite{hole,banks,emp}. The fact that the fundamental Planck scale is
as low as TeV also opens up an interesting possibility of
producing a large number of black holes at collider experiments
(e.g. LHC) \cite{scott,greg}.  Reference \cite{emp} showed that a
BH localized on a brane will radiate mainly in the brane, instead
of radiating into the Kaluza-Klein states of gravitons of the
bulk. In this case, the BH so produced will decay mainly into the
SM particles, which can then be detected in the detector.  This
opportunity has enabled investigation of the properties of BH at
terrestrial collider experiments.  There have been a number of
such studies
\cite{scott,greg,emparan,hoss,giddings,me1,me2,casa-lhc,hofmann,park,greg2,giudice,blei,solo,rizzo,RS-b,uhe-lhc} at
hadronic colliders.  A typical signature of the BH decay is a high
multiplicity, isothermal event, very much like a spherical
``fireball".
On the other hand, BH production has also been studied
in cosmic ray experiments or neutrino telescopes
\cite{kay,feng,anch,ratt,ring,uhe,kowalski,han,tu}.

Another interesting trans-Planckian phenomenon is the $p$-brane
($p$-B) \cite{ahn,jain,feng-p}. A BH can be considered as a
$0$-brane.  In particle collisions, if one considers BH
production, one should also consider $p$-brane production.  In
fact, the properties of $p$-branes reduce to those of BH in the
limit $p\to 0$. In extra dimension models, in which there are
large extra dimensions and small extra dimensions of the size of
Planck length, let a $p$-brane wraps on $r$ small and $p-r$ large
dimensions. It was found \cite{ahn} that the production of
$p$-branes is comparable to BH's only when $r=p$, i.e., the
$p$-brane wraps entirely  on the small dimensions only.  If $r<
p$, the production of $p$-branes would be suppressed by powers of
$(M_*/M_{\rm Pl})$, where $M_*$ is the fundamental scale of the
$4+n$ dimensions and $M_{\rm Pl}$ is the 4-dimensional Planck
scale. The decay of $p$-branes is not well understood. One
interesting possibility is cascade into branes of lower dimensions
until they reach the dimension of zero. Whether the zero brane is
stable depends on the model. Another possibility is the decay into
brane and bulk particles, thus experimentally the decay can be
observed. Or it can be a combination of cascade into
lower-dimensional branes and direct decays.

A naive picture of BH or $p$-brane production is as follows.
Hadronic production of BH's or $p$-branes at colliders is expected
when the colliding partons with a center-of-mass energy
$\sqrt{\hat s} \agt M_{\rm BH}$ and an impact parameter less than
$R_{\rm BH}$ or $R_{p \rm B}$. Here $R_{\rm BH}$ is the Schwarzschild
radius of a BH with mass $M_{\rm BH}$ and $R_{p\rm B}$ is the radius of the
$p$-brane.  This semi-classical argument calls for a
geometric approximation for the cross section for producing a BH
of mass $M_{\rm BH}$ as
\[
\sigma(M_{\rm BH} ) \approx \pi R_{\rm BH}^2 \;.
\]
This is valid if the colliding partons are confined to a 3-brane (the SM
brane.)

There are other scenarios such as fat branes \cite{fat} or universal extra
dimensions \cite{universal},
in which the SM particles are allowed, to some extent,
to propagate in the space dimensions other than the normal $3+1$
dimensions. \footnote{In these scenarios, there are $\sim
{\rm TeV}^{-1}$ size bounds on the dimensions that the SM particles can
propagate \cite{universal,gauge}. There are at the same time
some dimensions of very large size in order to bring the
fundamental Planck scale down to TeV.}
In the fat brane scenario, the original SM brane (with an
infinitesimal thickness) is extended into a small
but finite thickness (called a fat brane) in the extra dimension
coordinates \cite{fat}. The SM fermions are localized in various
locations on the fat brane in order that the Yukawa coupling can
be interpreted as the overlap of the fermion wave-functions
involved.  Interesting phenomenology in flavor physics arises in
this scenario.
The scenario of universal extra dimensions \cite{universal}
allows all SM particles
to move in the entire extra dimensions, excluding those dimensions
that are very large of mm size.  The size of these extra
dimensions that the SM can propagate can be as large as
$(300\,{\rm GeV})^{-1}$, depending on whether the lightest KK
state is stable or not.  Since the SM particles can move in the
extra dimensions, the conservation of momentum in the extra
dimensions turns into the conservation of the KK number.  The
phenomenology includes: (i) the KK states ($\pm n$) must be
pair-produced, and (ii) the lightest KK state is stable (this is rather
similar in spirit to the $R_p$ parity conservation in the supersymmetry.)

We mentioned these two scenarios because we want to point out that
if the SM particles also move in the extra dimensions, the cross
section for producing black holes or $p$-branes will be modified,
especially, the ratio of $p$-brane production to BH
production will be enhanced much more significantly than the usual $3+1$
dimensions.  This is the main result of the present work.
We shall use ``confined scenario'' to stand for the scenario in which
the SM particles are restricted to the SM brane, while
``unconfined scenario'' to stand for the scenario when the SM
particles are allowed to propagate in the extra dimensions.

The purpose of this work is to investigate the enhancement
in the ratio of $p$-brane cross section to BH cross section when the SM
particles also propagate in some of the extra dimensions. We show
that the ratio increases from $1-3.8$ to $1-105$ for $n=7$ (the
SM particles are allowed to propagate up to five dimensions of the $n=7$
extra dimensions, i.e., $p=m\le 5$.)
Thus, $p$-brane production is more interesting
in these fat brane and universal extra dimension scenarios.

\section{Scattering in $D>3+1$ dimensions}

In $D=3+1$ dimensions, the production cross section of BH's is
given by
\begin{equation}
\label{3dsigma}
 \sigma_{3+1}^{\rm BH}(M_{\rm BH} ) = \pi R_{\rm BH}^2 ,\;
\end{equation}
where $R_{\rm BH}$ is the Schwarzchild radius of the BH. This
formula is based on a geometric argument.  When two incoming
partons with a center-of-mass energy $\sqrt{\hat s} \ge M_{\rm
BH}$ collide each other at an impact parameter less than $R_{\rm
BH}$, a BH is formed.  Thus, the production cross section is given
by the above formula.

In the scenario that the SM particles also propagate in the extra
dimensions, the incoming partons must be colliding at an impact
parameter less than $R_{\rm BH}$ in all dimensions, in order to
produce a BH.  This is understood as follows.  In $D=3+1$
dimensions, the distance between the incoming partons in the extra
dimension coordinates is simply zero.  In $D>3+1$ dimensions, if
the scattering distance between the incoming partons are larger
than the $R_{\rm BH}$, they cannot merge into a BH.  Therefore,
the cross section for producing a BH must scale as $R_{\rm
BH}^{k+2}$, where $k$ is the number of extra dimensions that the
SM particles also propagate.  We therefore have
\begin{equation}
\label{kdsigma}
 \sigma_{3+1+k}^{\rm BH}(M_{\rm BH} ) = A_{4+k}\; R_{\rm BH}^{2+k} \;.
\end{equation}
Here $k$ is not necessarily equal to $n$, the total number of
extra dimensions. $A_{4+k}$ is a geometrical factor in $3+1+k$
dimension. In fact, in the following we shall introduce some small
extra dimensions of size $M_D^{-1}$ and some large extra
dimensions of mm size.  It is the large extra dimensions that
bring the fundamental Planck scale to TeV. The SM particles
propagate in the small extra dimensions but not the large extra
dimensions. Therefore, $k \le n-2$, where experimentally the
number of large extra dimensions must be at least two.

\subsection{Space time configuration}

Let there be $n$ total extra dimensions with $m$ small extra dimensions
and $n-m$ large extra dimensions.  When we say small extra dimensions, we
mean the size is of order of $1/M_*$, the fundamental Planck scale.
The observed 4D Planck scale $M_{\rm Pl}$ is then a derived quantity given by
\cite{arkani}
\begin{equation}
\label{nm}
M_{\rm Pl}^2 = M_*^{2+n} \, V_{m} \, V_{n-m} \;,
\end{equation}
where $V_{m}$ and $V_{n-m}$ are the volumes of the extra $m$ and $n-m$
dimensions, respectively, given by
\begin{equation}
V_m = L_m^m \equiv \left( \frac{ l_m}{M_*}\right)^m\,; \qquad
V_{n-m} = L_{n-m}^{n-m} \equiv \left( \frac{ l_{n-m}}{M_*}\right)^{n-m}\,,
\end{equation}
where we have expressed the lengths $L_m, L_{n-m}$ in units of Planckian length
$1/M_*$.

Suppose the small extra dimension has the size of $L_m \sim 1/M_*$, i.e.,
$l_m \sim 1$ then
\begin{equation}
\label{n}
M_{\rm Pl}^2 = M_*^{2} \; \left( l_{n-m}  \right)^{n-m} \;.
\end{equation}
The fundamental Planck scale $M_*$ is lowered to the TeV range if the size
$L_{n-m}$ is taken to be very large, of order $O({\rm mm})$.

In literature, another conventionally used definition of the fundamental
Planck scale $M_D$ is related to $M_*$ by
\begin{equation} \label{m*-md}
M_D^{n+2} = \frac{( 2\pi )^{n} }{ 8 \pi G_{4+n}} =
              \frac{( 2\pi )^{n} }{ 8 \pi} \, M^{n+2}_* \;,
\end{equation}
where $G_{4+n}$ is the gravitational constant in $D=4+n$ dimensions
(used in the Einstein equation:
${\cal R}_{AB} - \frac{1}{2} g_{AB} {\cal R} = -8\pi G_{4+n} T_{AB}$.)

\subsection{Black hole production}

A black hole is characterized by its mass, angular momentum, and charge.
Here we consider only the uncharged and non-rotating case.
The Schwarzschild radius $R_{\rm BH}$ of a BH of mass $M_{\rm BH}$ in
$4+n$ dimensions is given by \cite{myers}
\begin{equation}
\label{r} R_{\rm BH}(M_{\rm BH}) = \frac{1}{M_D}\; \left (
\frac{M_{\rm BH}}{M_D} \right)^{\frac{1}{n+1}}\; \left( \frac{ 2^n
\pi^{ \frac{n-3}{2}} \Gamma(\frac{n+3}{2} )}{n+2} \right
)^{\frac{1}{n+1}} \;,
\end{equation}
which is much smaller than the size of the extra dimensions.
Another important quantity that characterizes a BH is its entropy given by
\cite{myers}
\begin{equation}
\label{entropy} S_{\rm BH}(M_{\rm BH}) = \frac{4\pi}{n+2}\; \left
( \frac{M_{\rm BH}}{M_D} \right)^{\frac{n+2}{n+1}}\; \left( \frac{
2^n \pi^{ \frac{n-3}{2}} \Gamma(\frac{n+3}{2} )}{n+2} \right
)^{\frac{1}{n+1}} \;.
\end{equation}
To ensure the validity of the above
classical description of BH \cite{giddings,me1,me2}, the entropy must be
sufficiently large, of order 25 or so.
 In Ref. \cite{giddings,me1,me2} it was shown that when
$M_{\rm BH}/M_D \agt 5$, the entropy $S_{\rm BH}\agt 25$.
Therefore, to avoid getting into the nonperturbative regime of the
BH and to ensure the semi-classical validity, we restrict the mass
of the BH to be $M_{\rm BH} \ge  5 M_D$.
BH production is expected when the colliding partons with a
center-of-mass energy $\sqrt{\hat s} \agt M_{\rm BH}$ pass within
a distance less than $R_{\rm BH}$.
A black hole of mass $M_{\rm
BH}$ is formed and the rest of energy, if there is, is radiated as
ordinary SM particles. This semi-classical argument calls for a
geometric approximation for the cross section for producing a BH
of mass $M_{\rm BH}$ as in Eq. (\ref{3dsigma}).  In the case that
the SM particles also move in $k$ extra dimensions, the production
cross section is modified to Eq. (\ref{kdsigma}).

\subsection{$p$-brane production}

$p$-branes can also be formed in particle collisions;  in particular,
when there exist small extra dimensions of the size $\sim 1/M_*$ in addition
to the large ones of the size $\gg 1/M_*$.  It was pointed out
by Ahn {\it et al.} \cite{ahn} that the production cross section
of a $p$-brane completely wrapped on the small extra dimensions is larger than
that of a spherically symmetric black hole.

Consider an uncharged and static $p$-brane with a mass $M_{p \rm B}$ in $(4+n)$
dimensional space-time ($m$ small Planckian size and $n-m$ large size extra
dimensions such that $n \ge p$.)
Suppose the $p$-brane wraps on $r (\le m)$ small extra dimensions and on
$p-r (\le n-m)$ large extra dimensions.
Then the ``radius'' of the $p$-brane is
\begin{equation}
\label{rp} R_{p \rm B}(M_{p\rm B}) = \frac{1}{\sqrt{\pi} M_*} \,
\gamma(n,p) \, V_{p\rm B}^{ \frac{-1}{1+n-p} } \, \left(
\frac{M_{p\rm B}}{M_*} \right)^{ \frac{1}{1+n-p} } \;,
\end{equation}
where $V_{p\rm B}$ is the volume wrapped by the $p$-brane in units of the
Planckian length.
Recall from Eq. (\ref{nm}), $M_{\rm Pl}^2 = M_*^2 l_{n-m}^{n-m} l_m^m$,
where $l_{n-m}\equiv L_{n-m}\,M_{*}$ and $l_m \equiv L_m\,M_{*}$
are the lengths of the size of the large and small
extra dimensions in units of Planckian length ($\sim 1/M_*$).  Then
$V_{p\rm B}$ is given by
\begin{equation}\label{vp}
V_{p \rm B} = l_{n-m}^{p-r} \, l_m^{r} \approx \left( \frac{M_{\rm Pl}}{M_*}
\right )^{ \frac{2(p-r)}{n-m} } \;,
\end{equation}
where we have taken $l_m \equiv L_m \,M_{*} \sim 1$.
The function $\gamma(n,p)$ is given by
\begin{equation}\label{gamma}
\gamma(n,p) = \left[ 8 \Gamma \left( \frac{3+n-p}{2} \right) \sqrt{
\frac{1+p}{(n+2)(2+n-p)} } \right ]^{\frac{1}{1+n-p} } \;.
\end{equation}
The $R_{p\rm B}$ reduces to the $R_{\rm BH}$ in the limit $p=0$.

The production cross section of $p$-brane is similar to that of
BH's, based on a naive geometric argument \cite{ahn}. When the
partons collide with a center-of-mass energy $\sqrt{\hat s}$
larger than the fundamental Planck scale and an impact parameter
less than the size of the $p$-brane, a $p$-brane of mass $M_{p\rm
B} \le \sqrt{\hat s}$ can be formed.  That is
\begin{equation}
\label{3dpb}
 \hat \sigma_{3+1}^{p\rm B} (M_{p\rm B}) = \pi R^2_{p\rm B} \;,
\end{equation}
in the $D=3+1$ scattering.  In the scenario that the SM particles
also propagate in $k$ extra dimensions, the production cross
section for $p$-branes is modified to
\begin{equation}
\label{kdpb}
 \hat \sigma_{3+1+k}^{p\rm B} (M_{p\rm B}) = A_{4+k}\; F(s) \; R^{2+k}_{p\rm B}
 \;,
\end{equation}
where $F(s)$ is a dimensionless form factor of order one. For
simplicity we assume $F(s)=1$. Therefore, the production cross
section for $p$-brane is the same as BH's in the limit $p=0$
(i.e., a BH can be considered as a $0$-brane.)

\subsection{Ratio of $p$-brane to BH production}

In Eq. (\ref{rp}) we can see that the radius of a $p$-brane is
suppressed by some powers of the volume $V_{p\rm B}$ wrapped by
the $p$-brane.  It is then obvious that the production cross
section is largest when $V_{p\rm B}$ is minimal, in other words,
the $p$-brane wraps entirely on the small extra dimensions only,
i.e, $r=p$.  When $r=p$, $V_{p\rm B}=1$. We can also compare the
production cross section of $p$-branes with BH's. Assuming that
their masses are the same and the production threshold $M^{\rm
min}$ is the same, the ratio of cross sections in $D=3+1+k$
scattering is given by
\begin{equation}
\label{R} R \equiv \frac{\hat \sigma_{3+1+k}^{p\rm B} (M_{p\rm
B}=M)}{\hat \sigma_{3+1+k}^{\rm BH} (M_{\rm BH}=M)} = \left (
\frac{M_*}{M_{\rm Pl}} \right)^{\frac{2(2+k)(p-r)}{(n-m)(1+n-p)}}
\, \left(\frac{M}{M_*} \right )^{ \frac{(2+k)p}{(1+n)(1+n-p)} } \,
\left( \frac{\gamma(n,p)}{\gamma(n,0)} \right )^{2+k} \;,
\end{equation}
which reduces to the result in Ref. \cite{ahn} in the limit $k=0$.
In the above equation, the most severe suppression factor is in
the first parenthesis on the right hand side.  Since we are
considering a fundamental scale $M_* \sim$ TeV, the factor $(M_*/M_{\rm Pl})
\sim 10^{-16}-10^{-15}$. Thus, the only meaningful production of
$p$-brane occurs for $r=p$, and then their production is
comparable.  In table \ref{pbrane}, we show this ratio for various
values of $n$ and $p$.  We choose $k=m$, i.e., the SM particles
propagate in all small extra dimensions.  This is the maximal
choice of $k$.  We note that $p(=r) \le m$.  The maximal ratio
occurs if $p=m$.  We also choose the mass of BH and $p$-brane to
be $5 M_D$. Under these choices, the only free parameters are $n$
and $m$ (with $m \le n-2$.)

\section{Discussions}

Black hole and $p$-brane production only occurs at an energy scale
substantially larger than the fundamental Planck scale.  At this scale, the
SM particles already feel the presence and move in the small TeV$^{-1}$-sized
extra dimensions in the unconfined scenario.
Dimensional analysis tells us that the cross section scales
as $R^{2+k}$.  Therefore, the ratio of production of $p$-branes to
BH is now enhanced by a much larger factor than in the confined scenario.

One typical question to ask is: will the BH or $p$-brane decay into SM
particles such that it can be detected?  In the scenario that the SM particles
are confined to a 3-brane, it was shown \cite{emp} that the BH decays mostly
into the SM particles.  The naive argument is as follows.  The BH decays
via Hawking evaporation and the wavelength $\lambda$
of the thermal spectrum corresponding to
the Hawking temperature is much larger than the size of the BH.
Therefore, the BH radiates like a point source in $s$-waves such that
it decays equally into brane and bulk modes, and will not
see the higher angular momentum states available in the extra
dimensions.  Since on the brane there are many more particles than
in the bulk, and therefore the BH decays dominantly into brane modes.

The scenario considered in this work is different as the SM particles
are allowed to move off the SM brane when the energy scale is above a certain
scale ($\sim$ TeV).  However, we argue that when a BH is formed by colliding
two SM particles, the BH must be within a distance  $R_{\rm BH}$ away from
the SM brane.
Suppose two SM particles are accelerated on the SM brane, which is at
$y=0$ (here $y$ denotes collectively the coordinates of the extra dimensions.)
When the energy scale is above the compactification scale ($\sim$ TeV),
the SM particles begin to feel the extra dimensions and can move off the SM
brane.  However, for the two colliding particles to form a BH, the impact
distance between them  must be less than $R_{\rm BH}$.  We argue that the
chance that the two particles move far away from the SM brane to go to
nearly the same spot in the extra dimensions (within $R_{\rm BH}$ in each
dimension) is very tiny.
Therefore, when a BH is formed it is most likely to be within
$y=0 \pm R_{\rm BH}$, i.e., the BH is at least touching or
intersecting with the SM brane.
Hence, when the BH decays into SM particles
(each of only a few hundreds of GeV),
they should be observable on the SM brane.  A similar argument applies to
the $p$-brane formation and decay.

In this paper, we have pointed out that the production of $p$-branes
relative to black holes can be enhanced by a much larger factor in
the unconfined scenario that the SM particles are allowed to move in the extra
dimensions (e.g., fat-brane scenario and universal extra dimensions)
than in the confined scenario.
In the confined scenario,
the ratio of $p$-brane production to BH production can at most be
$3.8$ (for the case $n=7,m=5=p$), while 
in the unconfined case the ratio can be as large as $105$ (for
the case $n=7,m=5=p=k$).

\section*{Acknowledgments}
This research was supported in part by the National Center
for Theoretical Sciences under a grant from the National Science
Council of Taiwan R.O.C. 
C.-H. Chou was supported by the National Science Council of R.O.C. under
Grant Nos. NSC90-2811-M-001-042.


\newpage
\begin{table}[th!]
\caption{\small \label{pbrane} The ratio $R \equiv \hat
\sigma_{3+1+k}^{p\rm B} (M_{p\rm B}=M)/\hat \sigma_{3+1+k}^{\rm
BH} (M_{\rm BH}=M)$ of Eq. (\ref{R}) for various $n$ and $p$ with
$m\le n-2$ in the unconfined scenario (here $k$ stands for the
number of small extra dimensions that the SM particles can
propagate.) We have used $M_{\rm BH}=M_{p \rm B}=5 M_D$. We have
assumed that the $p$-brane wraps entirely on small extra
dimensions, i.e., $r=p$. In order to obtain the largest ratio $R$
we have chosen $p=m$ and $k=m$.  The numbers in the parenthesis
are for the confined scenario, i.e. $k=0$. }
\medskip
\begin{ruledtabular}
\begin{tabular}{ccccccc}
       & $p=0$ & $p=1$ & $p=2$ & $p=3$ & $p=4$ & $p=5$  \\
\hline
$n=2$ & 1 &  &&&&  \\
$n=3$ & 1 &  2.4(1.8) &&& &\\
$n=4$ & 1 &  1.7(1.4) & 6.1(2.5) && &\\
$n=5$ & 1 &  1.4(1.3) & 3.0(1.7) & 16 (3.0) & &\\
$n=6$ & 1 &  1.3(1.2) & 2.0(1.4) & 5.3(1.9) &41(3.5) & \\
$n=7$ & 1 &  1.2(1.1) & 1.6(1.3) & 2.9(1.5) & 9.3(2.1) & 105(3.8)\\
\end{tabular}
\end{ruledtabular}
\end{table}

\end{document}